\documentclass[prl,twocolumn,superscriptaddress,showpacs]{revtex4}

\usepackage{graphicx} % Include figure files
\usepackage{epsfig}   % Include figure files
\usepackage{amssymb,amsmath,amsfonts,hyperref}
\usepackage{wasysym}
\usepackage{latexsym}
\usepackage{comment}
\usepackage{color}

\newcommand{\la}{\langle}
\newcommand{\ra}{\rangle}

\newcommand{\nn}{\nonumber}

\newcommand{\nogr}[1]{{#1}}

\begin{document}

\title{Classical spin-liquid on the maximally frustrated honeycomb lattice}
\author{J. Rehn}
\affiliation{Max-Planck-Institut f\"ur Physik komplexer Systeme, 01187 Dresden, Germany}
\author{Arnab Sen}
\affiliation{Department of Theoretical Physics, Indian Association for the Cultivation of Science, Kolkata 700032, India}
\author{Kedar Damle}
\affiliation{Department of Theoretical Physics, Tata Institute of Fundamental Research, Mumbai 400 005, India}
\author{R. Moessner}
\affiliation{Max-Planck-Institut f\"ur Physik komplexer Systeme, 01187 Dresden, Germany}
\date{\today}

\begin{abstract}
We show that the honeycomb Heisenberg antiferromagnet with $J_1/2=J_2=J_3$,
where $J_{1/2/3}$ are first-, second- and third-neighbour couplings
respectively, forms a classical spin liquid with pinch-point singularities
in the structure factor at the Brillouin zone corners.
Upon dilution with non-magnetic ions, fractionalised degrees of freedom
carrying $1/3$ of the free moment emerge.
Their effective description in the limit of low-temperature is that of spins
randomly located on a triangular lattice, with a frustrated interaction of
long-ranged logarithmic form. The XY version of this magnet exhibits nematic
thermal order by disorder, which comes with a clear experimental diagnostic.
\end{abstract}

%\pacs{
%	xx
%}
\maketitle

%==============================================================================%
%\section{{\bf INTRODUCTION}}
{\it Motivation.}---The honeycomb lattice has -- somewhat belatedly -- become
one of the prime hunting grounds for spin liquids (SL) in
$d=2$~\cite{anderson1973resonating}, in addition to the kagome and the $J_1-J_2$
square lattice Heisenberg models, which have been the focus of much attention over
decades, continuing until today. In both these latter cases~\cite{chandra1988possible,
dagotto1989phase,figueirido1990exact,sachdev1992kagome,zhitomirsky1996valence,
singh2003symmetry,mambrini2006plaquette,richter2009the,yan2011spin,messio2012kagome},
confidence in the existence of a quantum SL state for
$S=1/2$ magnets has ebbed and flowed, while  the classical (large-spin)
versions evade liquidity by exhibiting --  rather interesting -- forms
of order by disorder~\cite{villain1980order,chandra1990ising,reimers1992order,
chalker1992hidden,ritchey1993spin,zhitomirsky2008octupolar,chern2012dipolar,
moessner1998low,moessner1998properties}. 

The richness of magnetic models on the honeycomb lattice -- bipartite, like the square
lattice -- has therefore come as somewhat of a surprise. Initially emulating its
brethren by appearing to support a quantum SL in a Hubbard model~\cite{meng2010quantum},
it has been attracting attention in the context of the fractionalised phases of the Kitaev
honeycomb model~\cite{kitaev2003fault}, exhibiting highly unusual exactly soluble
quantum SL phases. Particular impetus arose from the suggestion that the
Kitaev Hamiltonian may describe the materials ${\{Na,Li\}}_2IrO_3$,
provided a Heisenberg term is added~\cite{singh2012relevance,mazin2012na2iro3,jackeli2009mott}.

In fact, detailed studies of these materials suggest that further nearest
neighbor terms play an important role in explaining spiral ordering at
low temperatures~\cite{reuther2014spiral}, and one of the models studied in
some detail is the  $J_1-J_2-J_3$ Heisenberg model, which had already been
subject to considerable earlier
attention~\cite{fouet2001investigation,reuther2011magnetic,albuquerque2011phase,bishop2015frustrated}.
In determining the Hamiltonian appropriate to these materials, it has
turned out to be instructive to consider their response to disorder~\cite{andrade2014magnetism}.

Here, we identify and study in detail an unusual, hitherto
overlooked, classical SL state on the honeycomb lattice, associated
with the (known) degeneracy point $J_1/2=J_2=J_3$ of the Heisenberg model on
the honeycomb lattice. It exhibits remarkable new features. These arise from
the fact that the dual lattice, as well as the underlying Bravais lattice, is
the tripartite triangular lattice. They include pinch points in the structure
factor at the zone corner wavevector ${\bf Q}$ (which distinguishes between
the three sublattices), as well as novel disorder effects whereby, upon
dilution, fractionalised moments carrying {\it one third} of the microscopic
spin moment appear. These fractionalized moments interact via a frustrated,
sublattice-dependent, long range interaction in the limit of low temperature,
$T$.

\begin{figure}
\epsfig{file=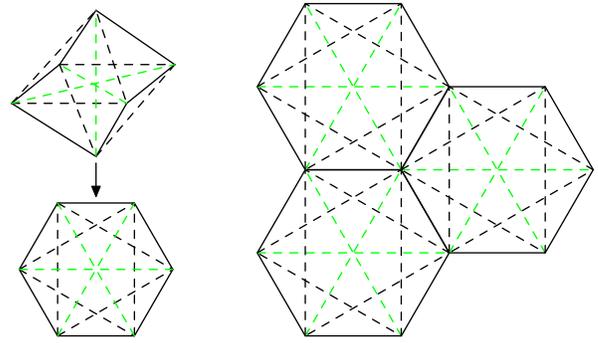,width=0.9\linewidth,angle=0}
\caption{(color online). Projection of the octahedron into the hexagon and the $J_1-J_2-J_3$ model
on the honeycomb lattice. The $J_3$ interactions are differentiated with colors.}
\label{fig:OctProj}
\end{figure}

This model is further remarkable as it can be thought of the first realisation of
a SL in $d=2$ of {\it edge-sharing} simplices, which here take the form
of octahedra. In addition, its XY version does exhibit nematic order by disorder,
which turns out to be straightforwardly detectable in neutron scattering through the
appearance of peaks in the structure factor.  

The remainder of this paper is organised as follows. We first introduce
the model and derive and describe its SL, for which we formulate a novel
low-energy description. We then study its behaviour under dilution.
All our analytical predictions are supported by Monte Carlo (MC) simulations of the
microscopic Hamiltonian. We close with an outlook, in which we argue that this
model is quite natural, as (i) the degeneracy point corresponds to a reasonably
natural set of parameter values; and (ii) we expect the SL to fan out
as $T$ is increased, at the expense of adjacent phases exhibiting lower entropies.

%==============================================================================%
%\section{{\bf MODEL STUDIED}}
{\it Model.}---The Hamiltonian for classical $O(n)$ spins $\vec{S}_i$ of unit
length on sites $i$ of the honeycomb lattice reads: 
\begin{align}
H &= J_1\sum_{\la i,j\ra}\vec{S}_i\cdot\vec{S}_j + J_2\sum_{\la\la i,k\ra\ra}\vec{S}_i\cdot\vec{S}_k
  + J_3\sum_{\la\la\la i,l\ra\ra\ra}\vec{S}_i\cdot\vec{S}_l \nn \\
  &= \frac{J}{2}\sum_{\alpha}(\vec{S}_{\hexagon}^\alpha)^2 + \text{const.},
  \label{eq:constr}
\end{align}
where $\la i,j\ra$, $\la\la i,k\ra\ra$ and $\la\la\la i,l\ra\ra\ra$ refer respectively
to first, second and third nearest neighbour pairs, while the second line follows from
fixing $J_1/2 = J_2 = J_3$. 

\begin{figure}
\nogr{\epsfig{file=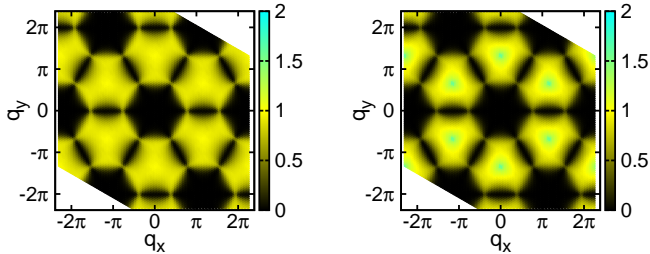,width=0.7\linewidth,angle=0}}
\caption{(color online). Structure factor as obtained in Monte Carlo simulations of the pristine
Heisenberg (left) and XY (right) systems. Both results correspond to $N=1800$ spins
at $T/J=0.01$.}
\label{fig:stFactor}
\end{figure}

This form shows that each and any configuration where each hexagon, labelled
by $\alpha$, has vanishing total spin, $\vec{S}_{\hexagon}^\alpha=0$, is a
ground state. Such a rewriting is often helpful for geometrically frustrated
lattices. It is most often used for `corner-sharing' structure of elementary
simplices~\cite{moessner1998properties,moessner1998low}, examples being
pyrochlore (corner-sharing tetrahedra) or kagome (corner-sharing triangles)
lattices. It immediately allows to estimate the dimensionality of the ground
state manifold, $F$. This proceeds by subtracting the number of constraints,
$K$, imposed by Eq.~\ref{eq:constr}, from the total number of degrees of
freedom, $D$, of the spin system.

For a system of $n$-component spins with $N$ such simplices, and each spin part of
$b$ simplices, $D=q(n-1)/b$ per simplex, where the number of spins in a simplex
$q=3,4,6$ for triangle, tetrahedron, octahedron respectively. Each simplex imposes
$K=n$ constraints, as each component of its total spin must vanish. Hence,
\begin{equation}
F= \frac{q(n-1)}{b} - n.
\label{eq:GSdim}
\end{equation} 
To maximize $F$, and hence enhance the chance of finding a
SL~\cite{moessner1998properties,moessner1998low}, one thus should
minimize $b$, or maximise $n$ and $q$. Indeed, $b$ is minimal for
corner-sharing arrangements, and $q=4$, $n=3$ result in the well-established
classical SL on the pyrochlore lattice.
Triangle-based lattices (kagome has $q=3$) need higher, $n\geq 4$, component
spins for a similar SL to arise~\cite{huse1992classical}.

The $J_1-J_2$ model on the square lattice with $J_2=J_1/2$ can be thought of as
edge-sharing tetrahedra, with a large $q=4$; it does not support
$F>0$ for any $n$. Indeed, no such Heisenberg model with $F>0$
has been identified for edge-sharing simplices at all so far.

However, from Eq.~\ref{eq:GSdim}, $F=1$ for $q=6$ and $b=3$, which corresponds
to the frustration point of the honeycomb lattice,
Eq.~\ref{eq:constr}! It can be thought of as edge-sharing {\it octahedra}
(Fig.~\ref{fig:OctProj}), and thus presents the first instance
of a possible SL on an edge-sharing lattice. It is also the first with
$b>2$, a fact with significant consequences as we explore in detail below.

Before we do this, we demonstrate that this model does indeed
exhibit a SL with algebraic correlations for $T\rightarrow0$,
by analytically evaluating the correlations for soft spins
(equivalently, in the large-$n$ or self-consistent Gaussian
approximation~\cite{garanin1999classical}) and comparing the result
to classical MC simulations. These yield a $T=0$ structure
factor presenting {\it pinch points}, the defining
characteristic of such algebraic SLs~\cite{henley2010coulomb}.
Somewhat unusually, in this case the pinch points are located at the corners
of the Brillouin zone.

\begin{figure}
\nogr{\epsfig{file=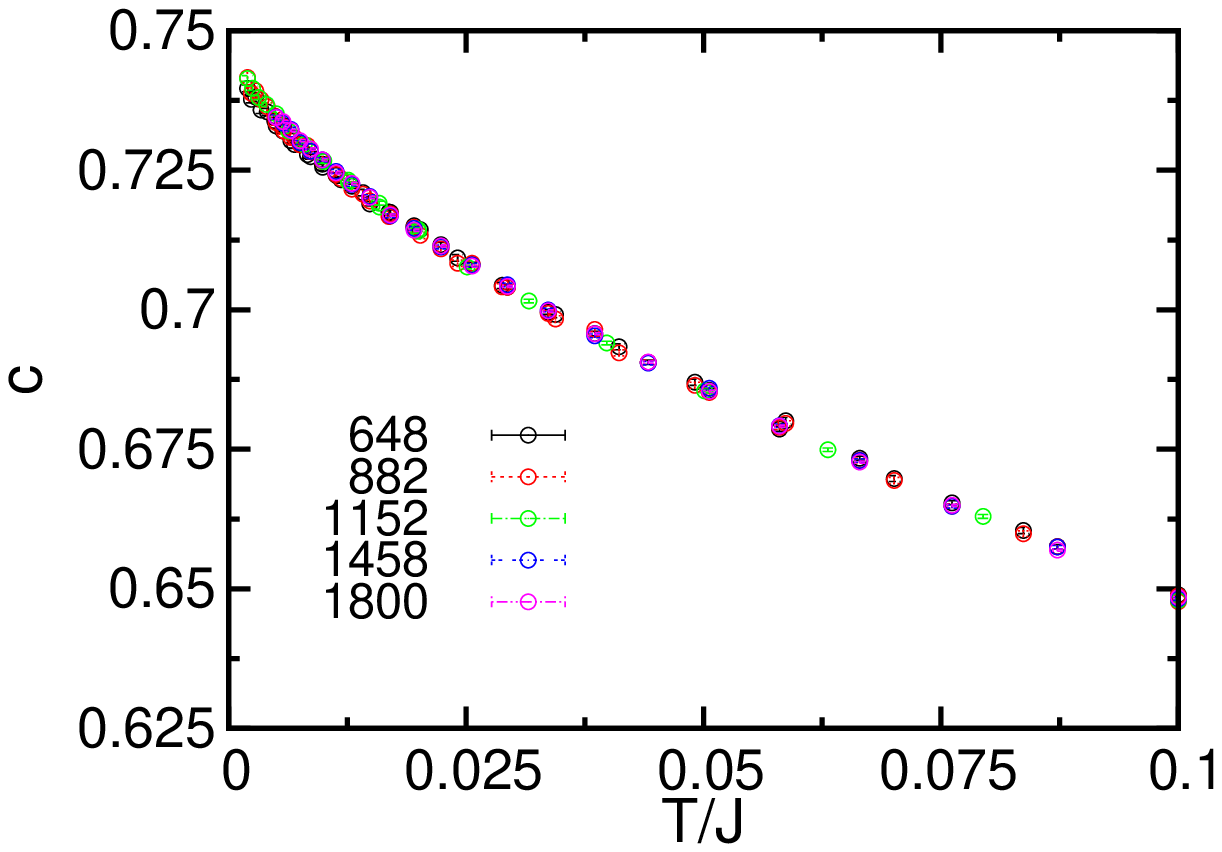,width=\linewidth,angle=0}}
\nogr{\epsfig{file=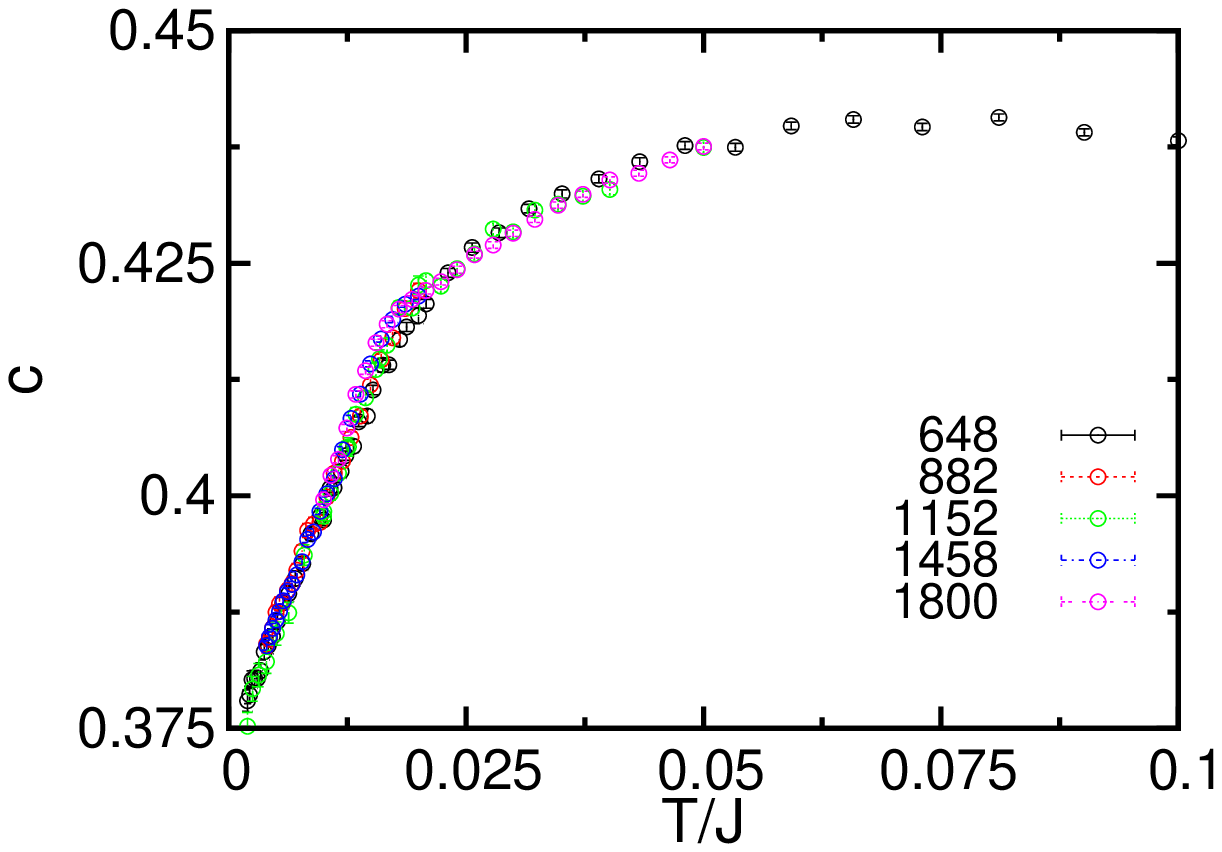,width=\linewidth,angle=0}}
\caption{(color online). Specific heat obtained in Monte Carlo simulations of the pristine
Heisenberg ($n=3$, top) and XY ($n=2$, bottom) systems, with
$c=\frac{3}{8}k_B < \frac{bn}{2q}k_B$ indicating nematic order by disorder
for the XY case.}
\label{fig:spHeat}
\end{figure}

From our MC simulations for Heisenberg and XY spins, we plot 
structure factor and specific heat on Figs.~\ref{fig:stFactor}, \ref{fig:spHeat}.
The MC simulations employ a combination of heat-bath and microcanonical moves
as well as parallel tempering moves.
The structure factor from MC simulation of Heisenberg spins agrees with the
analytical prediction for soft spins.
By contrast, for $n=2$, the corresponding XY model, low temperature peaks
develop in addition to the pinch points. This is an instance of nematic
(collinear) order by disorder, as is readily verified by constraint
counting~\cite{moessner1998properties,moessner1998low}.
We note in passing that the appearance of these peaks provides an
unusually direct signature of collinear ordering.

This interpretation is confirmed by a low-$T$ specific heat of $c=0.375k_B$
per spin, reduced from the value of $c=bnk_B/2q$ expected from equipartition
in the absence of order by disorder~\cite{chalker1992hidden,ritchey1993spin,zhitomirsky2008octupolar,chern2012dipolar,moessner1998low},
as is found in the Heisenberg magnet with $c=0.75 k_B$.

The existence of pinch points in the Heisenberg case comes as somewhat
of a surprise given the non-bipartite nature of the dual triangular lattice.
In the corresponding corner-sharing models, the bipartiteness of the dual
lattice (square, honeycomb or diamond lattice) is a crucial ingredient
for such pinch points~\cite{henley2010coulomb}. Indeed, in work close in spirit
to the present one, on bosons on a honeycomb and the dual triangular
lattice~\cite{motrunich2003bosonic}, one finds an Ising emergent gauge
field implying the absence of pinch points. The way this issue resolves itself
in the present case is quite interesting: First, the pinch points are
located at the Brillouin zone corners, corresponding to a three-sublattice
wave vector ${\bf Q}$.
Second, the low-energy description is naturally expressed in terms of
a vector field that captures slow modulations near wavevector-${\bf Q}$,
reminiscent of the two dimensional height field acting as an emergent
$U(1)$ gauge field~\cite{henley2010coulomb}. In detail, this proceeds
as follows.

Consider an A-sublattice (B-sublattice) site $\vec{r}_A$ ($\vec{r}_B$)
of the honeycomb lattice, which sits at the center of an
``up-pointing'' (``down-pointing'') triangle comprising
dual lattice points $\vec{R}_a$, $\vec{R}_b$ and $\vec{R}_c$ belonging
to the three sublattices of the tripartite dual triangular lattice.
One writes the corresponding $O(n)$ spins $\vec{S}_{\vec{r}}$ in terms
of $\vec{\zeta}_{\vec{R}}$ and $\vec{\tau}_{\vec{R}}$, two $O(n)$ vector
fields on the dual triangular lattice.
\begin{align}
\vec{S}_{\vec{r}_A} = \sum_{\alpha=a,b,c}(\vec{\tau}_{\vec{R}_{\alpha}} + \vec{\zeta}_{\vec{R}_{\alpha}}),
\hspace{4mm}
\vec{S}_{\vec{r}_B} = \sum_{\alpha=a,b,c}(\vec{\tau}_{\vec{R}_{\alpha}}  - \vec{\zeta}_{\vec{R}_{\alpha}}). \nonumber
\end{align}
In the self-consistent Gaussian approximation, the partition function
for the Hamiltonian Eq.~\ref{eq:constr} can be written as a product
of $\vec{\zeta}$ and $\vec{\tau}$ partition functions, with actions
\begin{align}
{\mathcal{S}}_{\zeta} = F_1(\{\vec{\zeta}\}),
\hspace{4mm}
{\mathcal{S}}_{\tau} = F_1(\{\vec{\tau}\}) + F_2(\{\vec{\tau}\}),
\end{align}
where
\begin{eqnarray}
F_1(\{\vec{v}\}) = \frac{\rho}{2} \sum_{\vec{r}} (\vec{v}_{\vec{R}_a(\vec{r})} + \vec{v}_{\vec{R}_b(\vec{r})} + \vec{v}_{\vec{R}_c(\vec{r})})^2 
\end{eqnarray}
for $\{\vec{v}\} = \{\vec{\zeta}\}$ or $\{\vec{\tau}\}$, and 
\begin{align}
F_2(\{\vec{\tau}\}) = \frac{\beta J}{2} \sum_{\vec{R}} (6\vec{\tau}_{\vec{R}} + 2\sum_{\vec{R}_n \in \partial\vec{R}} \vec{\tau}_{\vec{R}_n})^2
\end{align}
here, $\vec{R}_n \in \partial\vec{R}$ denotes the six dual triangular
lattice sites $\vec{R}_n$ that are nearest neighbours of the dual
triangular lattice site $\vec{R}$. The stiffness constant $\rho$ is
adjusted to yield $\langle \vec{S}_{\vec{r}}^2 \rangle = 1$.

This action implies that $\vec{\zeta}$ encodes the $T=0$
fluctuations of the classical SL, while $\vec{\tau}$ captures thermal
fluctuations. The $T\rightarrow0$ limit is thus characterized by a
particularly simple action in which the $\vec{\tau}$ fields do not contribute.
This action, as well as the expressions for the physical spins $\vec{S}$,
are both invariant under
$\vec{\zeta}(\vec{R}) \rightarrow \vec{\zeta}(\vec{R}) +
 {\rm Re} (\vec{\chi} \exp(2\pi i {\mathbf Q} \cdot \vec{R}))$
for any $\vec{\chi}$.

\iffalse
\begin{figure}
\nogr{\epsfig{file=HexLattFlux-3.eps,width=\linewidth,angle=0}}
\caption{The emergent gauge fields live on the bonds of the red, green and blue
honeycomb lattices.}
\label{fig:LatticeFlux}
\end{figure}
\fi
%==============================================================================%
%\section{{\bf DILUTION}}

{\it Dilution Effects.}---The ground states of SLs often are
less revealing of their topological nature than their excitations. An
elegant way to visualise the latter as effectively a ground state property
is to introduce disorder which then nucleates excitations. In SLs, this is
perhaps most easily done by replacing some of the magnetic ions with
non-magnetic ones. For classical SLs, this dilution problem has been studied
in some detail both experimentally~\cite{obradors1988magnetic,schiffer1997two,
mendels2000ga,laforge2013quasispin},
and theoretically~\cite{henley2001effective,moessner1999magnetic,sen2011fractional,sen2012vacancy}.
In particular, for the cases of SCGO, the checkerboard and the pyrochlore
lattices, it was found that fractional impurity moments carrying one half
of the moment of a free spin arise as a cooperative phenomenon. These so-called
orphan spins occur when all but one of the spins of a simplex are replaced
-- so that the total spin of that simplex (see Eq.~\ref{eq:constr})
can no longer possibly vanish. 

\begin{figure}
\nogr{\epsfig{file=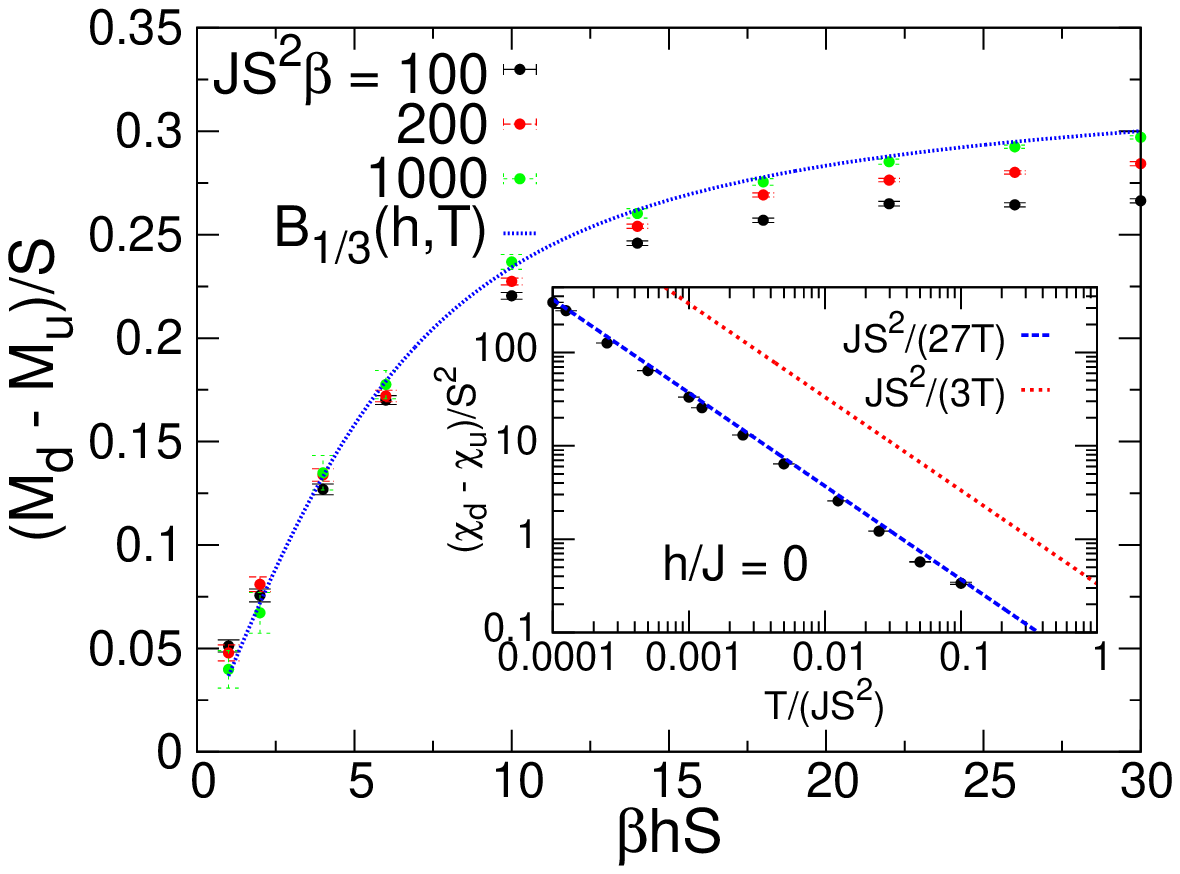,width=\linewidth,angle=0}
{\epsfig{file=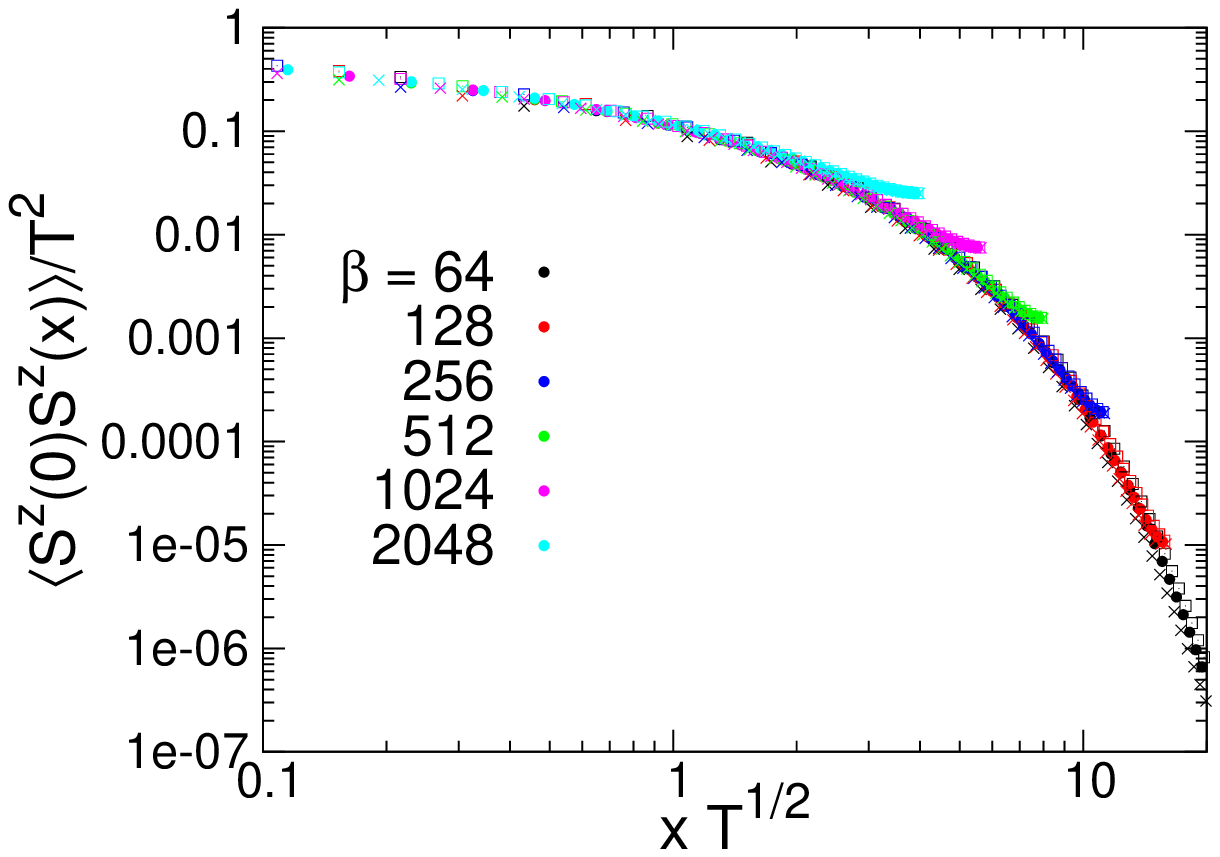,width=\linewidth,angle=0}}
\put(-233,66.8){\rotatebox{90}{\hexagon}}
\put(-233,94.3){\rotatebox{90}{\hexagon}}}
\caption{(color online). {\bf Top:} `Impurity magnetization', defined as the difference of
total magnetization in the diluted and undiluted systems, as observed in MC
simulations of the model.
The solid curve corresponds to the theoretical prediction for a free spin
$S/3$ in a field $h$, i.e., the Langevin function $\mathcal{B}_{S/3}(h,T)$.
The inset shows the `impurity susceptibility' at zero external field,
consistent with a Curie law for fractionalized spins $S/3$.
{\bf Bottom:} Testing the scaling prediction for the charge correlations
on a finite lattice of linear size $L=210$ using the expression for
correlations within the soft spin approach.
Crucially, correlations between sites on the same sublattice have been
multiplied by an extra scaling factor of $-1/2$.}
\label{fig:DiluEff}
\end{figure}

These orphans turn out to provide a number of signatures of the new structure
of the honeycomb SL. First of all, they directly reflect the fact that we have
$b=3$ {\it edge-sharing} octahedra meeting in each site -- the fractional impurity
moment is not one half but {\it one third} of that of a free spin! This is displayed
in Fig.~\ref{fig:DiluEff} (top panel) where a calculation based on a hybrid hard-soft
spin theory~\cite{sen2011fractional,sen2012vacancy} is compared with numerical
results for the local susceptibility. This is, to our knowledge, the first
instance of fractionalisation into three items in a classical spin model.

Interactions between these orphans are entropic in nature and take the form
of an effective Heisenberg exchange $J_{\text{eff}}$.
They are mediated by the bulk SL, and hence reflect the structure
of the latter. In the classical SLs known so far, these effective
interactions can be written in a form which is uniformly
antiferromagnetic~\cite{rehn2015random}. Here, this is not possible:
We now find that these interactions are antiferromagnetic / ferromagnetic
for orphans residing on the same / different sublattice of the dual
triangular lattice, respectively, with the antiferromagnetic interactions
being twice as strong as the ferromagnetic ones. This intricate structure
in the effective exchange couplings follows from our field theory, which
relates these entropic interactions to a bulk property of the pristine
spin liquid, namely the correlations between the thermally excited net
spins $\vec{S}_{\hexagon}$ (Eq.~\ref{eq:constr}): 
\begin{align}
\beta J_{\text{eff}} \approx
\frac{-\la\vec{S}_{\hexagon,\vec{r}_1}\cdot\vec{S}_{\hexagon,\vec{r}_2}\ra}{\la\vec{S}_{\hexagon,\vec{r}} \cdot \vec{S}_{\hexagon,\vec{r}} \ra^2} ,
\end{align}

For low $T$ and large distances $|\vec{r}_1-\vec{r}_2|\gg a$, where $a$ is
the lattice spacing, this gives a scaling form:
\begin{align}
\beta J_{\text{eff}} &= \eta(\vec{r}_1,\vec{r}_2) {\mathcal{F}}((\vec{r}_1-\vec{r}_2)\sqrt{T}) \\
 &\overset{T\rightarrow0}{=} \frac{1}{2\pi}\eta(\vec{r}_1,\vec{r}_2)\log(\vec{r}_1-\vec{r}_2) ,
\end{align}
where $\eta=+1$ ($\eta=-1/2$) if the orphans are on the same (different)
sublattices of the dual triangular lattice.
This is verified using the analytical large-n result for a finite lattice,
Fig.~\ref{fig:DiluEff}. In the limit $T\rightarrow0$, $\beta J_{\text{eff}}$
exhibits a long-ranged logarithmic form.

%==============================================================================%

%==============================================================================%
%\section{{\bf CONCLUSIONS}}
{\it Outlook.}---Our model, notwithstanding its simplicity, displays a plethora
of phenomena of current interest; the unusual emergent $\vec{\tau}$ fields and the new
fractionalized behavior of $1/3$ for the impurity spin moments show that even in a
classical setting, these nontrivial phenomena, up to now apparently constrained to
the quantum realm, can emerge. The particular new frustrated logarithmic interactions
between the impurity moments are as yet unstudied, and will possibly lead to a spin glass,
unlike in the bipartite cases~\cite{rehn2015random}.

As for realisations, the $2:1:1$ ratio of exchange interactions is natural
if exchange is via an ion on the hexagon center with no angular dependence,
as the nearest neighbors bonds are part of two hexagons. Known experimental
values are encouragingly nearby, being close to $2:1.6:1.6$~\cite{mazin2012na2iro3}.
Hence direct observation of these phenomena might be possible,
the main obstacle perhaps being finite quantum fluctuations for $S=1/2$. Quite
generally, at finite $T$, the classical SL behavior will be favoured
over competing phases on account of its large entropy, and in particular fan out
from the degeneracy point. We hope that this work will incite further investigation
on appropriate honeycomb materials.

\section*{Acknowledgements:}
The work of AS is partly supported through the Partner Group program between the
Indian Association for the Cultivation of Science (IACS) and the Max Planck Institute
for the Physics of Complex Systems. JR acknowledges the hospitality of IACS during
the initial stages of this work. This work was in part supported by DFG via
SFB 1143. We thank G. Jackeli, A. L\"auchli and J. Richter for discussions.

%==============================================================================%
\bibliography{ref_nonbipCoul}

\begin{thebibliography}{44}
\expandafter\ifx\csname natexlab\endcsname\relax\def\natexlab#1{#1}\fi
\expandafter\ifx\csname bibnamefont\endcsname\relax
  \def\bibnamefont#1{#1}\fi
\expandafter\ifx\csname bibfnamefont\endcsname\relax
  \def\bibfnamefont#1{#1}\fi
\expandafter\ifx\csname citenamefont\endcsname\relax
  \def\citenamefont#1{#1}\fi
\expandafter\ifx\csname url\endcsname\relax
  \def\url#1{\texttt{#1}}\fi
\expandafter\ifx\csname urlprefix\endcsname\relax\def\urlprefix{URL }\fi
\providecommand{\bibinfo}[2]{#2}
\providecommand{\eprint}[2][]{\url{#2}}

\bibitem[{\citenamefont{{P.~W.~Anderson}}(1973)}]{anderson1973resonating}
\bibinfo{author}{\bibnamefont{{P.~W.~Anderson}}},
  \bibinfo{journal}{Mat.~Res.~Bulletin} \textbf{\bibinfo{volume}{8}},
  \bibinfo{pages}{153} (\bibinfo{year}{1973}).

\bibitem[{\citenamefont{{P. Chandra and B.
  Doucot}}(1988)}]{chandra1988possible}
\bibinfo{author}{\bibnamefont{{P. Chandra and B. Doucot}}},
  \bibinfo{journal}{Phys. Rev. B} \textbf{\bibinfo{volume}{38}},
  \bibinfo{pages}{9335(R)} (\bibinfo{year}{1988}).

\bibitem[{\citenamefont{{Elbio Dagotto and Adriana
  Moreo}}(1989)}]{dagotto1989phase}
\bibinfo{author}{\bibnamefont{{Elbio Dagotto and Adriana Moreo}}},
  \bibinfo{journal}{Phys. Rev. Lett.} \textbf{\bibinfo{volume}{63}},
  \bibinfo{pages}{2148} (\bibinfo{year}{1989}).

\bibitem[{\citenamefont{{F. Figueirido, A. Karlhede, S. Kivelson, S. Sondhi, M.
  Rocek, and D. S. Rokhsar}}(1990)}]{figueirido1990exact}
\bibinfo{author}{\bibnamefont{{F. Figueirido, A. Karlhede, S. Kivelson, S.
  Sondhi, M. Rocek, and D. S. Rokhsar}}}, \bibinfo{journal}{Phys. Rev. B}
  \textbf{\bibinfo{volume}{41}}, \bibinfo{pages}{4619} (\bibinfo{year}{1990}).

\bibitem[{\citenamefont{{Subir Sachdev}}(1992)}]{sachdev1992kagome}
\bibinfo{author}{\bibnamefont{{Subir Sachdev}}}, \bibinfo{journal}{Phys. Rev.
  B} \textbf{\bibinfo{volume}{45}}, \bibinfo{pages}{12377}
  (\bibinfo{year}{1992}).

\bibitem[{\citenamefont{{M. E. Zhitomirsky and Kazuo
  Ueda}}(1996)}]{zhitomirsky1996valence}
\bibinfo{author}{\bibnamefont{{M. E. Zhitomirsky and Kazuo Ueda}}},
  \bibinfo{journal}{Phys. Rev. B} \textbf{\bibinfo{volume}{54}},
  \bibinfo{pages}{9007} (\bibinfo{year}{1996}).

\bibitem[{\citenamefont{{R. R. P. Singh, W. Zheng, J. Oitmaa, O. P. Sushkov,
  and C. J. Hamer}}(2003)}]{singh2003symmetry}
\bibinfo{author}{\bibnamefont{{R. R. P. Singh, W. Zheng, J. Oitmaa, O. P.
  Sushkov, and C. J. Hamer}}}, \bibinfo{journal}{Phys. Rev. Lett.}
  \textbf{\bibinfo{volume}{91}}, \bibinfo{pages}{017201}
  (\bibinfo{year}{2003}).

\bibitem[{\citenamefont{{Matthieu Mambrini, Andreas L\"auchli, Didier
  Poilblanc, and Fr\'ed\'eric Mila}}(2006)}]{mambrini2006plaquette}
\bibinfo{author}{\bibnamefont{{Matthieu Mambrini, Andreas L\"auchli, Didier
  Poilblanc, and Fr\'ed\'eric Mila}}}, \bibinfo{journal}{Phys. Rev. B}
  \textbf{\bibinfo{volume}{74}}, \bibinfo{pages}{144422}
  (\bibinfo{year}{2006}).

\bibitem[{\citenamefont{{J. Richter and J.
  Schulenburg}}(2010)}]{richter2009the}
\bibinfo{author}{\bibnamefont{{J. Richter and J. Schulenburg}}},
  \bibinfo{journal}{Eur. Phys. J. B} \textbf{\bibinfo{volume}{73}},
  \bibinfo{pages}{117} (\bibinfo{year}{2010}).

\bibitem[{\citenamefont{{Simeng Yan, David A. Huse, Steven R.
  White}}(2011)}]{yan2011spin}
\bibinfo{author}{\bibnamefont{{Simeng Yan, David A. Huse, Steven R. White}}},
  \bibinfo{journal}{Science} \textbf{\bibinfo{volume}{332}},
  \bibinfo{pages}{1173} (\bibinfo{year}{2011}).

\bibitem[{\citenamefont{{Laura Messio, Bernard Bernu, and Claire
  Lhuillier}}(2012)}]{messio2012kagome}
\bibinfo{author}{\bibnamefont{{Laura Messio, Bernard Bernu, and Claire
  Lhuillier}}}, \bibinfo{journal}{Phys. Rev. Lett.}
  \textbf{\bibinfo{volume}{108}}, \bibinfo{pages}{207204}
  (\bibinfo{year}{2012}).

\bibitem[{\citenamefont{{J. Villain, R. Bidaux, J.-P. Carton and R.
  Conte}}(1980)}]{villain1980order}
\bibinfo{author}{\bibnamefont{{J. Villain, R. Bidaux, J.-P. Carton and R.
  Conte}}}, \bibinfo{journal}{J. Physique} \textbf{\bibinfo{volume}{41}},
  \bibinfo{pages}{1263} (\bibinfo{year}{1980}).

\bibitem[{\citenamefont{{P. Chandra, P. Coleman, and A. I.
  Larkin}}(1990)}]{chandra1990ising}
\bibinfo{author}{\bibnamefont{{P. Chandra, P. Coleman, and A. I. Larkin}}},
  \bibinfo{journal}{Phys. Rev. Lett.} \textbf{\bibinfo{volume}{64}},
  \bibinfo{pages}{88} (\bibinfo{year}{1990}).

\bibitem[{\citenamefont{{J. N. Reimers and A. J.
  Berlinsky}}(1993)}]{reimers1992order}
\bibinfo{author}{\bibnamefont{{J. N. Reimers and A. J. Berlinsky}}},
  \bibinfo{journal}{Phys. Rev. B} \textbf{\bibinfo{volume}{48}},
  \bibinfo{pages}{9539} (\bibinfo{year}{1993}).

\bibitem[{\citenamefont{{J. T. Chalker, P. C. W. Holdsworth, and E. F.
  Shender}}(1992)}]{chalker1992hidden}
\bibinfo{author}{\bibnamefont{{J. T. Chalker, P. C. W. Holdsworth, and E. F.
  Shender}}}, \bibinfo{journal}{Phys. Rev. Lett.}
  \textbf{\bibinfo{volume}{68}}, \bibinfo{pages}{855} (\bibinfo{year}{1992}).

\bibitem[{\citenamefont{{I. Ritchey, P. Chandra and P.
  Coleman}}(1993)}]{ritchey1993spin}
\bibinfo{author}{\bibnamefont{{I. Ritchey, P. Chandra and P. Coleman}}},
  \bibinfo{journal}{Phys. Rev. B} \textbf{\bibinfo{volume}{47}},
  \bibinfo{pages}{15342(R)} (\bibinfo{year}{1993}).

\bibitem[{\citenamefont{{M. E. Zhitomirsky}}(2008)}]{zhitomirsky2008octupolar}
\bibinfo{author}{\bibnamefont{{M. E. Zhitomirsky}}}, \bibinfo{journal}{Phys.
  Rev. B} \textbf{\bibinfo{volume}{78}}, \bibinfo{pages}{094423}
  (\bibinfo{year}{2008}).

\bibitem[{\citenamefont{{Gia-Wei Chern and R.
  Moessner}}(2013)}]{chern2012dipolar}
\bibinfo{author}{\bibnamefont{{Gia-Wei Chern and R. Moessner}}},
  \bibinfo{journal}{Phys. Rev. Lett.} \textbf{\bibinfo{volume}{110}},
  \bibinfo{pages}{077201} (\bibinfo{year}{2013}).

\bibitem[{\citenamefont{{R. Moessner and J. T.
  Chalker}}(1998{\natexlab{a}})}]{moessner1998low}
\bibinfo{author}{\bibnamefont{{R. Moessner and J. T. Chalker}}},
  \bibinfo{journal}{Phys. Rev. B} \textbf{\bibinfo{volume}{58}},
  \bibinfo{pages}{12049} (\bibinfo{year}{1998}{\natexlab{a}}).

\bibitem[{\citenamefont{{R. Moessner and J. T.
  Chalker}}(1998{\natexlab{b}})}]{moessner1998properties}
\bibinfo{author}{\bibnamefont{{R. Moessner and J. T. Chalker}}},
  \bibinfo{journal}{Phys. Rev. Lett.} \textbf{\bibinfo{volume}{80}},
  \bibinfo{pages}{2929} (\bibinfo{year}{1998}{\natexlab{b}}).

\bibitem[{\citenamefont{{Z. Y. Meng, T. C. Lang, S. Wessel, F. F. Assaad and A.
  Muramatsu }}(2010)}]{meng2010quantum}
\bibinfo{author}{\bibnamefont{{Z. Y. Meng, T. C. Lang, S. Wessel, F. F. Assaad
  and A. Muramatsu }}}, \bibinfo{journal}{Nature}
  \textbf{\bibinfo{volume}{464}}, \bibinfo{pages}{847} (\bibinfo{year}{2010}).

\bibitem[{\citenamefont{{A.~Yu.~Kitaev}}(2003)}]{kitaev2003fault}
\bibinfo{author}{\bibnamefont{{A.~Yu.~Kitaev}}}, \bibinfo{journal}{Annals of
  Physics} \textbf{\bibinfo{volume}{303}}, \bibinfo{pages}{2}
  (\bibinfo{year}{2003}).

\bibitem[{\citenamefont{{Yogesh Singh, S. Manni, J. Reuther, T. Berlijn, R.
  Thomale, W. Ku, S. Trebst, and P. Gegenwart}}(2012)}]{singh2012relevance}
\bibinfo{author}{\bibnamefont{{Yogesh Singh, S. Manni, J. Reuther, T. Berlijn,
  R. Thomale, W. Ku, S. Trebst, and P. Gegenwart}}}, \bibinfo{journal}{Phys.
  Rev. Lett.} \textbf{\bibinfo{volume}{108}}, \bibinfo{pages}{127203}
  (\bibinfo{year}{2012}).

\bibitem[{\citenamefont{{I. I. Mazin, Harald O. Jeschke, Kateryna Foyevtsova,
  Roser Valent\'i, and D. I. Khomskii}}(2012)}]{mazin2012na2iro3}
\bibinfo{author}{\bibnamefont{{I. I. Mazin, Harald O. Jeschke, Kateryna
  Foyevtsova, Roser Valent\'i, and D. I. Khomskii}}}, \bibinfo{journal}{Phys.
  Rev. Lett.} \textbf{\bibinfo{volume}{109}}, \bibinfo{pages}{197201}
  (\bibinfo{year}{2012}).

\bibitem[{\citenamefont{{G. Jackeli and G.
  Khaliullin}}(2009)}]{jackeli2009mott}
\bibinfo{author}{\bibnamefont{{G. Jackeli and G. Khaliullin}}},
  \bibinfo{journal}{Phys. Rev. Lett.} \textbf{\bibinfo{volume}{102}},
  \bibinfo{pages}{017205} (\bibinfo{year}{2009}).

\bibitem[{\citenamefont{{Johannes Reuther, Ronny Thomale, and Stephan
  Rachel}}(2014)}]{reuther2014spiral}
\bibinfo{author}{\bibnamefont{{Johannes Reuther, Ronny Thomale, and Stephan
  Rachel}}}, \bibinfo{journal}{Phys. Rev. B} \textbf{\bibinfo{volume}{90}},
  \bibinfo{pages}{100405(R)} (\bibinfo{year}{2014}).

\bibitem[{\citenamefont{{J. B. Fouet, P. Sindzingre, C.
  Lhuillier}}(2001)}]{fouet2001investigation}
\bibinfo{author}{\bibnamefont{{J. B. Fouet, P. Sindzingre, C. Lhuillier}}},
  \bibinfo{journal}{Eur. Phys. J. B} \textbf{\bibinfo{volume}{20}},
  \bibinfo{pages}{241} (\bibinfo{year}{2001}).

\bibitem[{\citenamefont{{Johannes Reuther, Dmitry A. Abanin, and Ronny
  Thomale}}(2011)}]{reuther2011magnetic}
\bibinfo{author}{\bibnamefont{{Johannes Reuther, Dmitry A. Abanin, and Ronny
  Thomale}}}, \bibinfo{journal}{Phys. Rev. B} \textbf{\bibinfo{volume}{84}},
  \bibinfo{pages}{014417} (\bibinfo{year}{2011}).

\bibitem[{\citenamefont{{A. F. Albuquerque, D. Schwandt, B. Het\'enyi, S.
  Capponi, M. Mambrini, and A. M. L\"auchli}}(2011)}]{albuquerque2011phase}
\bibinfo{author}{\bibnamefont{{A. F. Albuquerque, D. Schwandt, B. Het\'enyi, S.
  Capponi, M. Mambrini, and A. M. L\"auchli}}}, \bibinfo{journal}{Phys. Rev. B}
  \textbf{\bibinfo{volume}{84}}, \bibinfo{pages}{024406}
  (\bibinfo{year}{2011}).

\bibitem[{\citenamefont{{R. F. Bishop, P. H. Y. Li, O. G\"otze, J. Richter, and
  C. E. Campbell}}(2015)}]{bishop2015frustrated}
\bibinfo{author}{\bibnamefont{{R. F. Bishop, P. H. Y. Li, O. G\"otze, J.
  Richter, and C. E. Campbell}}}, \bibinfo{journal}{arxiv:1504.02275}
  (\bibinfo{year}{2015}).

\bibitem[{\citenamefont{{Eric C. Andrade and Matthias
  Vojta}}(2014)}]{andrade2014magnetism}
\bibinfo{author}{\bibnamefont{{Eric C. Andrade and Matthias Vojta}}},
  \bibinfo{journal}{Phys. Rev. B} \textbf{\bibinfo{volume}{90}},
  \bibinfo{pages}{205112} (\bibinfo{year}{2014}).

\bibitem[{\citenamefont{{David A. Huse and Andrew D.
  Rutenberg}}(1992)}]{huse1992classical}
\bibinfo{author}{\bibnamefont{{David A. Huse and Andrew D. Rutenberg}}},
  \bibinfo{journal}{Phys. Rev. B} \textbf{\bibinfo{volume}{45}},
  \bibinfo{pages}{7536(R)} (\bibinfo{year}{1992}).

\bibitem[{\citenamefont{{D. A. Garanin and Benjamin
  Canals}}(1999)}]{garanin1999classical}
\bibinfo{author}{\bibnamefont{{D. A. Garanin and Benjamin Canals}}},
  \bibinfo{journal}{Phys. Rev. B} \textbf{\bibinfo{volume}{59}},
  \bibinfo{pages}{443} (\bibinfo{year}{1999}).

\bibitem[{\citenamefont{{C. L. Henley}}(2010)}]{henley2010coulomb}
\bibinfo{author}{\bibnamefont{{C. L. Henley}}}, \bibinfo{journal}{Annu. Rev.
  Condens. Matter Phys.} \textbf{\bibinfo{volume}{1}}, \bibinfo{pages}{179}
  (\bibinfo{year}{2010}).

\bibitem[{\citenamefont{{O. I. Motrunich}}(2003)}]{motrunich2003bosonic}
\bibinfo{author}{\bibnamefont{{O. I. Motrunich}}}, \bibinfo{journal}{Phys. Rev.
  B} \textbf{\bibinfo{volume}{67}}, \bibinfo{pages}{115108}
  (\bibinfo{year}{2003}).

\bibitem[{\citenamefont{{X. Obradors, A. Labarta, A. Isalgu\'e, J. Tejada, J.
  Rodriguez, and M. Pernet}}(1988)}]{obradors1988magnetic}
\bibinfo{author}{\bibnamefont{{X. Obradors, A. Labarta, A. Isalgu\'e, J.
  Tejada, J. Rodriguez, and M. Pernet}}}, \bibinfo{journal}{Sol. State Commun.}
  \textbf{\bibinfo{volume}{65}}, \bibinfo{pages}{189} (\bibinfo{year}{1988}).

\bibitem[{\citenamefont{{P. Schiffer and I. Daruka}}(1997)}]{schiffer1997two}
\bibinfo{author}{\bibnamefont{{P. Schiffer and I. Daruka}}},
  \bibinfo{journal}{Phys. Rev. B} \textbf{\bibinfo{volume}{56}},
  \bibinfo{pages}{13712} (\bibinfo{year}{1997}).

\bibitem[{\citenamefont{{P. Mendels, A. Keren, L. Limot, M. Mekata, G. Collin,
  and M. Horvati\'c}}(2000)}]{mendels2000ga}
\bibinfo{author}{\bibnamefont{{P. Mendels, A. Keren, L. Limot, M. Mekata, G.
  Collin, and M. Horvati\'c}}}, \bibinfo{journal}{Phys. Rev. Lett.}
  \textbf{\bibinfo{volume}{85}}, \bibinfo{pages}{3496} (\bibinfo{year}{2000}).

\bibitem[{\citenamefont{{A. D. LaForge, S. H. Pulido, R. J. Cava, B. C. Chan,
  and A. P. Ramirez}}(2013)}]{laforge2013quasispin}
\bibinfo{author}{\bibnamefont{{A. D. LaForge, S. H. Pulido, R. J. Cava, B. C.
  Chan, and A. P. Ramirez}}}, \bibinfo{journal}{Phys. Rev. Lett.}
  \textbf{\bibinfo{volume}{110}}, \bibinfo{pages}{017203}
  (\bibinfo{year}{2013}).

\bibitem[{\citenamefont{{C. L. Henley}}(2001)}]{henley2001effective}
\bibinfo{author}{\bibnamefont{{C. L. Henley}}}, \bibinfo{journal}{Can. J.
  Phys.} \textbf{\bibinfo{volume}{79}}, \bibinfo{pages}{1307}
  (\bibinfo{year}{2001}).

\bibitem[{\citenamefont{{R. Moessner and A. J.
  Berlinsky}}(1999)}]{moessner1999magnetic}
\bibinfo{author}{\bibnamefont{{R. Moessner and A. J. Berlinsky}}},
  \bibinfo{journal}{Phys. Rev. Lett.} \textbf{\bibinfo{volume}{83}},
  \bibinfo{pages}{3293} (\bibinfo{year}{1999}).

\bibitem[{\citenamefont{{Arnab Sen, Kedar Damle, and R.
  Moessner}}(2011)}]{sen2011fractional}
\bibinfo{author}{\bibnamefont{{Arnab Sen, Kedar Damle, and R. Moessner}}},
  \bibinfo{journal}{Phys. Rev. Lett.} \textbf{\bibinfo{volume}{106}},
  \bibinfo{pages}{127203} (\bibinfo{year}{2011}).

\bibitem[{\citenamefont{{Arnab Sen, Kedar Damle, and R.
  Moessner}}(2012)}]{sen2012vacancy}
\bibinfo{author}{\bibnamefont{{Arnab Sen, Kedar Damle, and R. Moessner}}},
  \bibinfo{journal}{Phys. Rev. B} \textbf{\bibinfo{volume}{86}},
  \bibinfo{pages}{205134} (\bibinfo{year}{2012}).

\bibitem[{\citenamefont{{J. Rehn, Arnab Sen, Alexei Andreanov, Kedar Damle, R.
  Moessner, and A. Scardicchio}}(2015)}]{rehn2015random}
\bibinfo{author}{\bibnamefont{{J. Rehn, Arnab Sen, Alexei Andreanov, Kedar
  Damle, R. Moessner, and A. Scardicchio}}}, \bibinfo{journal}{Phys. Rev. B}
  \textbf{\bibinfo{volume}{92}}, \bibinfo{pages}{085144}
  (\bibinfo{year}{2015}).

\end{thebibliography}
%==============================================================================%
\end{document}